**Solid-like rheological response of non-entangled polymers in the molten state**

H. Mendil, P. Baroni, L. Noirez*

Laboratoire Léon Brillouin (CEA-CNRS), Ce-Saclay, 91191 Gif-sur-Yvette France

**Abstract.** We show that under non-slippage conditions, non-entangled polymers display an elastic-like behaviour at a macroscopic scale up to at least hundred degrees above the glass transition temperature. This result observed on side-chain liquid crystal polymers and on ordinary polymers, is in contradiction with the flow behaviour usually described.

The measurements are carried out with a conventional rheometer at several tenths millimetre sample thicknesses, thus probing bulk properties. This elasticity implies that the melt contains a macroscopic cohesion and that collective motions and time scales longer than the individual polymer dynamics are involved. The identification of this solid-like property of molten non-entangled polymers is of considerable importance for a better understanding of the polymer dynamics.

**1. Introduction.**

Processing is the destiny of fluids materials. Controlling and understanding their flow behaviour is of broad scientific and technological interest in field daily used as extrusion processes, wetting, adhesion or lubrification.

An example of impressive effects of the shear flow is the recent discovery of a shear induced phase transition in melts of side-chain liquid crystal polymers (LC-polymers) [1]. These long time scale non-linear phenomena are explained neither by a flow coupling with the conventional viscoelastic relaxation times [2] nor with phase pretransitional dynamics [3]. We reveal that polymers are macroscopically solid-like even far above any phase transition and suspect these solid-like polymer properties to be at the origin of the non-linear effects. We show that this new elastic character is generic to ordinary polymers contradicting the conventional description in terms of flow behaviour. First observations of solid-like behaviours in polymers have been identified at molecularly thin (10-40nm) films [4] and are related to confinement properties. To our knowledge, the first work reporting on frozen times at larger scales, is a piezorheometer study, carried out by Martinoty and coworkers who identified an unexpected elastic response up to 50µm at 109°C and 45°C respectively above the glass transition in LC-Polymers and Polystyrene melts [5,6]. It was proposed that the melt presents dynamic macroscopic heterogeneities (elastic clusters). Dynamic heterogeneities have been reported using NMR [7] and Dynamic Light Scattering experiments [8] on glass forming materials revealing spatial extensions of 10nm and $10^3$Å respectively. These clusters are theoretically explained in terms either of coexistence of transient orientational and positional order [9] or of long range correlation of density fluctuations [10] frozen at the time scale of the measurements [5,6].

We show in this paper that we are able to access to a macroscopic (tenths millimetre scale) elastic response using a conventional rheometer and controlling the material wetting properties. We reveal this property in polymers melts of different chemical compositions and molecular weights.

Our demonstration is illustrated with the dynamic relaxation spectrum displayed by a LC-polymer melt at various temperatures and by an ordinary non-entangled polymer melt, measured with optimised polymer/substrate interactions. The paper is organised as followed:
We first study the macroscopic dynamic response of a methyl-phenyl benzyl substituted polyacrylate (PAOCH$_3$) far above the glass transition temperature. This LC-polymer is composed of a polymer chain onto the side of which mesogenic molecules have been grafted. Liquid crystal moieties assemble in a long range ordering (mesophases) up to the isotropic phase (liquid state) and display a strong ability to anchor on various substrates [11]. We then investigate the rheological response of an ordinary non-entangled polymer at different temperatures above the glass transition (at least at T-Tg=+89°C). To exclude any surface induced crystallisation, the polymer chosen is amorphous. The candidate is a poly(n-butyl acrylate) sample (PBuA).

## 2. Experimental.

Table1 summarises the characteristics of the two polymers.

*Table 1: \* The symbols I, N and $S_A$ correspond to the isotropic, nematic and smectic phases respectively. PD stands for Polymerisation Degree.*

|  | *PAOCH$_3$* | *PBuA* |
|---|---|---|
| *Chemical formula* | 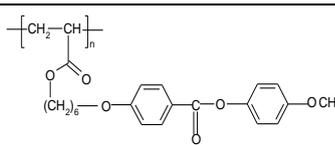 *Mw=30200, Mw/Mn=2.3 (PD=75)* | 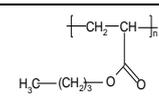 *Mw=20000, Mw/Mn=1.1 (PD=160)* |
| *Phase diagram* | *Tg - 22°C - $S_A$ - 89°C – N - 116°C - I\** | *Tg = -64 °C Amorphous* |
| *Working temperatures* | *100 – 160°C* | *25°C – 50°C* |
| *Sheared thickness l* | *0.025mm to 0.300mm* | *0.020mm up to 0.100mm* |

The rheological measurements are carried out in the dynamic mode in strain controlled conditions with an ARES rheometer equipped with an air-pulsed oven. This thermal environment ensures temperature control within 0.1°C. The samples were placed between plate-plate (diameter 12mm and 10mm) or cone-plate (diameter 20mm and angle 2.25°) surface treated fixtures. The zero gap is set by contact, the thicknesses are thus minima; the error is estimated of +0.010mm with respect to the indicated value. Since the experimental conditions are crucial, the samples are heated far above the glass transition (80 to 100°C), submitted to vacuum and shear to get rid of micro-bubbles and thermalised during several days before starting measurements. The molecular weight was controlled before and after the rheological investigation.

## 3. Study of the Liquid Crystal Polymer melt.
### 3.1. Identification of an elastic response in the Liquid Crystal Polymer melt.

Fig.1a and 1b display the frequency dependence of $G'(\omega)$ and $G''(\omega)$ in the liquid (isotropic) state, fourteen degrees over the Isotropic-Nematic transition ($T$=130°C), of the LC-polymer PAOCH$_3$ sheared at a thickness of 0.300mm. The storage modulus exhibits a non-trivial behaviour. Instead of a classical flow behaviour at low frequencies, the invariance of *G'* with respect to the frequency defines an elastic plateau of magnitude $G_p'$ (value averaged on the plateau frequency interval).

At low strain rate below a critical value ($\gamma \leq \gamma_{NL}=1\%$), $G_p'$ is independent of the strain defining a linear strain response regime (Fig.1c). This non-disturbed value is estimated of about $10^3$Pa at a 0.300mm gap thickness. In this linear regime at low frequency, the viscous modulus *G"* is independent of the frequency; its contribution is negligible with respect to $G_p'$ that merges with the complex modulus *G\** value; the system is solid-like. This elastic behaviour revealed here on low molecular weight non-entangled polymers is comparable to the rubbery plateau of high molecular weight entangled polymers [2]. This result contradicts the conventional description of the polymer dynamics, we will see that it can be explained using non-slip considerations and that it is also coherent with previous piezorheometry measurements [5].

### 3.2. Analysis of the non-linear behaviour:

As shown in Fig.1a, above a low critical strain $\gamma_{NL}$, the elastic plateau becomes strain dependant: the larger the strain, the lower is the elastic plateau $G_p'$. An additional mechanism superposes to the initial linear elastic response, giving rise to a dissipative regime; i.e. the non-linear regime.

Fig.1c gathers the strain dependence of $G_p'$ at different thicknesses at T=130°C (T-$T_{NI}$=+14°C). It shows that the non-linear regime is all the more rapidly reached that the thickness and the strain is large. Both large strains and large thicknesses correspond actually to conditions of high polymer displacement amplitude to the substrate; i.e. the non-linearity signs the entrance in a slippage regime.

The slippage process can be better evidenced by reporting the stress $\sigma=G_p'.\gamma$ measured by the transducer as a function of the applied strain amplitude (Fig.1d). When the system enters in the slippage regime, the energy excess is dissipated in the slip process. It results in an apparent decrease of the elastic modulus ($G'=\sigma/\gamma$) since the stress is no more proportional to the strain (non-linear regime), the excess being evacuated by dissipative slippage. Fig.1d shows that at low thickness $\sigma$ increases with the strain, whereas at larger thickness (>0.1mm) a maximum is reached above the critical strain $\gamma_{NL}$.

When $\sigma$ saturates (stress plateau), the highest stress that can accept the system is reached; it is a yield value (estimated of about $10^3$Pa) which measures the anchoring strengths. It is also associated to a critical length scale: $l_c=\gamma_{NL}.l$ (of the order of 3000nm for $l$=0.300mm) which describes the ultimate deformation length before entering in the slippage process; i.e. measuring the degree of "slippiness" of the elastic property. When $\gamma>\gamma_{NL}$, the elastic part of the material subtracts itself from the excess strain via a slippage process. Since slipping is a dissipative process, it gives rise to the non-linearity. The elasticity is kept but not properly (conservatively) measurable. It is interesting to notice that both linear and non-linear responses probe the elastic character of the material; the sliding ability being also a solid-like property.

### 3.3. Influence of the thickness on the linear elastic behaviour

We have seen (Fig.1c) that the strain dependence (non-linear regime) of the elastic plateau is all the more pronounced that the thickness is large. Let's focus on the elasticity thickness dependence measured under non-dissipative conditions; the linear elastic response.

The inset of Fig.1c displays the evolution of $G_p'$ measured in the linear regime, at $\gamma$=0.5%, as a function of the thickness. $G_p'$ is constant for thicknesses ranging from 0.025 up to 0.075mm. Above this critical thickness ($l^*$=0.075mm) $G_p'$ decreases very rapidly, approximately as a power law. We attribute the thickness induced softening by the high strain aptitude, which characterises weakly solid systems ($G_p'$ is weak compared to standard solids ($10^6$-$10^7$ Pa)). This thickness dependence indicates also that a critical gap, and thus a characteristic length scale, may exist. Similarly, thickness dependences are also observed in the piezorheometry measurements and are interpreted in terms of size-dependent elastic clusters [5].

### 3.4 Influence of the temperature on the linear elastic behaviour

Fig.1e illustrates the evolution of $G_p'(T)$ versus temperature at 0.300mm gap thickness and $\gamma$=0.5%. The temperature interval ranges over 60°C. Temperatures over 160°C were not tempted because of a possible thermal degradation of the acrylate ester [12]. The elastic plateau is not clearly sensitive to the temperature and is independent of the Isotropic-Nematic transition [5] ($T_{NI}$=116±0.5°C). It means that the elasticity does not originate from mesomorphic properties and is not reminiscent from phase pretransitional dynamics. It is a melt property already verified in other LC-polymers [13] and that should be observable on ordinary polymers. Our observations are completely convergent with the elastic behaviour described using the piezorheometer set-up [5].

**4. Study of the ordinary polymer melt.**

**4.1. Wetting effect and ordinary polymer behaviour.**

We have seen that the elastic behaviour can be revealed providing that the non-slip condition is fulfilled [6]. Indeed, anchoring and small strain displacements are the necessarily conditions to obtain the genuine "solid-like" response. Under these conditions, the LC-polymer appears solid-like even hundred degrees above the glass transition. In the case of the LC-polymers, we suspect a particularly strong anchoring since the first contact to the solid surface is created via an interfacial liquid crystal layer [11]. Our experience shows that a long time thermalisation and the choice of adapted surfaces optimise the anchoring interaction to the surfaces [14].

To quantify the effect of the substrate on the wetting properties, we determine the evolution of the profile of a drop laid on adapted and on standard aluminium surfaces. Fig.2a illustrates the evolution of the dynamic contact angle θ of a drop in the molten state versus time on these two different surfaces (the polymer dropped is the above described PBuA). This contact angle provides a simple and efficient measurement of the interaction between the polymer and the wettability of the substrate [15]. The polymer flattens rapidly on the treated surface: θ decreases more rapidly and the fluid spreads totally to wet the substrate, the dynamic contact angle θ reaching a value close to zero [16].

The dynamic response obtained with treated surfaces makes an elastic plateau emerged, showing that in addition to small strain amplitudes, the nature of the interactions between the substrate and the sample is important. In the treated surfaces, we suppose that the interactions are of electrostatic or of acido-basic type. We consider that the equilibrium conditions are reached when this elastic response does not evolve anymore. These effects are illustrated on Fig.2b (a similar evolution is displayed by the LC-polymer). To exclude any experimental artefact, the transducer response is measured in the same conditions (strain amplitude, frequency range, and temperature) without sample (low modulus curve in Fig.2b). There are more than 2 orders of magnitude between the elastic plateau level of the PBuA and the background noise, excluding any instrumental artefact contribution.

In contrast, using standard Aluminium fixtures, the conventional behaviour is recovered. Fig.2c displays the master curve constructed by time-temperature superposition at $T_{ref}$=25°C with the same polymer (PBuA - Table1). The absence of rubbery plateau at high frequencies confirms that the melt is non-entangled ($M_e$=22000g.mol$^{-1}$ for a butylacrylate backbone [17] and higher for LC-polymers [18]). The conventional viscoelastic relaxation time can be determined by the intersection of $G''(\omega)$ and $G'(\omega)$ fitting with $\omega$ and $\omega^2$ scaling respectively. It is estimated of about $4.10^{-4}$s at T=25°C (corresponding to 90°C above the glass transition temperature).

We have seen that when the melt interacts strongly with the surface fixtures, the storage modulus does not fit anymore with $\omega^2$ scaling (flow behaviour) but it displays a frequency plateau $G_p'$ characterised by a low value dispersion and a higher modulus (Fig.2d). This unsettled result implies that the conventional flow behaviour is observed under slippage conditions. Such surprising conclusion is nevertheless coherent with various wetting and dewetting experiments. The reproducibility and the frequent use of preshear in conventional experiments could suggest that the flow behaviour is actually linked to autophobocity conditions [19]. The reinforcement of the modulus values signs an enhancement of the anchoring conditions. The dynamic response of the PBuA obtained as a function of the strain rate in these conditions is illustrated on Fig.2d.

Fig.2e illustrates the influence of the sample thickness on $G_p'$ versus strain rate at T=25°C, showing that the elastic response is slightly dissipative at large thicknesses and may reveal a saturation at low thickness, as displayed in the inset of Fig.2e, consistent with what was already observed for the LC-polymer of this paper (the critical thickness is about $l^*$=0.075mm in the case of the LC-polymer against $l^*$=0.030mm for PBuA) and a LC-polysiloxane studied by piezorheometry (critical thicknesses of 0.015mm were reported [5]). It is interesting to notice that using different set-ups (conventional rheometer and piezorheometer) and different polymers, similar features are revealed. Comparing now the elastic moduli displayed by two ordinary non-entangled polymers (PBuA and the Polystyrene of reference [6]), the elastic behaviour of the PBuA and of the Polystyrene are revealed from 0.020mm up to 0.100mm, and from 0.015mm up to 0.050mm respectively. At a sample thickness of 0.020mm, the polystyrene elastic plateau is $G_0'$(T-Tg=+70°C)≈$10^4$Pa, whereas at a same gap thickness, the polybutylacrylate plateau is $G_0'$(T-Tg=+100°C)≈$10^3$Pa. These moduli are comparable; this elastic property is very likely a generic polymer blend characteristic.

**4.2. How conciliating the conventional and the elastic dynamic responses?**

Compared to the LC-polymer, the ordinary polymer presents an additional feature; both non-linear elastic plateau at low frequencies and classical viscoelastic behaviour are superposed at high frequency and high strain (Fig.2d).

The high strain rate tends thus to restore the conventional linear viscoelastic response at high strain and high frequency. The linear viscoelastic behaviour corresponds thus to the strongly slippage (non-linear) regime of the elastic response. The elastic modulus can be expressed using two independent modes:

-at $\gamma < \gamma_{NL}$: linear regime: $G' = G_0'(\ell) + \eta.\tau.\omega^2/(1 + \omega^2.\tau^2)$.

-at $\gamma > \gamma_{NL}$: non-linear regime: $G' = G_{NL}'(\ell) + \eta.\tau.\omega^2/(1 + \omega^2.\tau^2)$,

where $G_{NL}'(\ell) = G_0'(\ell).exp[-(\gamma - \gamma_{NL}/\gamma_{NL})]^\alpha$. $G_{NL}'$ is the non-linear elastic modulus; $G_0'$ and $\gamma_{NL}$ are the non-perturbed (linear) modulus and strain of the solid-like response. $\alpha$ is a dissipation exponent illustrating the decrease in modulus due to the sliding of the solid-like response. $\tau$ is the viscoelastic relaxation time.

At low frequency (long time scale) and low slippage (low strain), the first mode dominates. By increasing the strain, the elastic slippage increases making the classical viscoelastic response appeared. As in the case of the LC-polymer, a yield stress is reached at high strain, observable by the saturation of the stress $\sigma = G_p'.\gamma$ (Fig.2f). The comparison of the two polymers shows interestingly that the loss modulus, $G''$, is ω scaling whatever the strain in the case of the ordinary polymer whereas in the case of the LC-polymers, $G''$ displays a plateau-like behaviour of negligible moduli compared to the elastic component (Fig.1b). The LC-polymer system is completely frozen. But, the ordinary chain (PBuA) keeps a certain degree of mobility observable by the ω scaling behaviour of $G''$. Within our observation window, this behaviour change is a transition from gel-like state to solid-like state. It is very meaningful to interpret the origin of this unexpected elasticity. In the case of LC-polymers, the friction factor may be enhanced by the length of the side-branches and the short range polar interactions induced by the phenyl rings. Both effects play certainly a role in the giant elasticity observed. The long range correlation remains however to be elucidated.

Finally, in the case of the ordinary polymer, above T=75°C (T=Tg + 115°C), the elastic plateau can no more be accurately determined due to high dispersion in the measurements. It is interpreted by a temperature induced softening of the elastic response. This is consistent with long range glass transition effects as proposed in [6].

**5. Conclusions.**

In this paper we identify, using a conventional rheometer, an elastic behaviour at a macroscopic scale in non-entangled polymers at temperatures over hundred degrees above the glass transition. This is a strong and fundamental result in contradiction with the conventional description of the polymer dynamics but in complete agreement with first piezorheometry measurements.

We first show that LC-polymers present in the liquid (isotropic) state, far above phase and glass transition, a strong elastic response up to 0.300mm ($G_0' \cong 600$Pa). This is a novel type of elasticity, not due to chemical heterogeneities or a cross-linking and not originating from mesomorphic properties.

The study of the ordinary polymer (without mesomorphic graftings) confirms that the solid-like character is a polymer property. One fundamental consequence is that, far above phase and glass transitions, time scales longer than the conventional viscoelastic "terminal" time exist. The elastic character implies a connectivity between chains. Classical models of viscoelasticity (Rouse, reptation models) cannot account for cooperative molecular motions. In this regard, theoretical models involving mode coupling and relaxation theories [9, 20] seem to be more appropriate.

In the literature, solid-state properties of polymer melts have been identified at a molecular scale or a multiple of that [4], we scanned up to 0.300 millimetres and reveal still a strong elastic response. Certainly, at several molecular lengths, the dimensional restriction produces solidification effects. We report on much larger scales than confinement effects. In their pioneering work, Martinoty et al report already on a macroscopic elasticity. Using a piezorheometer, they identified either a low thickness linear elastic response below 0.050mm and 0.080mm for a Polystyrene chain and a LC-polymer respectively [5,6] or a large thickness flow behaviour response. From the comparis²on of the results obtained with two different techniques, using different substrates, with different polymers, at different temperatures away from transition temperatures, the same properties have been evidenced. These similarities demonstrate the fundamental elastic character of non-entangled polymers which are wrongly considered as flowing fluids above the glass transition temperature.

By analysing the non-linearity as originating from a molecular sliding motion over a macroscopic distance when the yield stress is exceeded, we take into account both responses passing continuously from the low static friction which reveals the native elastic plateau to a high dynamic friction state which shows the classical viscoelasticity. The "linear" regime of the viscoelasticity corresponds thus to non-linear conditions of the fundamental solid-like response. This elasticity is kept but not properly measurable (entrance in the non-linearity). The experiment conditions (surface energy, wetting properties, homogeneity of the melt…) are determinant; they fix the scale at which the medium disconnects from the linear response, i.e. when the slippage process takes place. Elastic response and slippage are the counter parts of a same cohesive property, observed here for the first time with a conventional rheometer in non-entangled melts.

Since both linear and non-linear responses reflect characteristic features of a solid-like behaviour, and this even far from any transition temperatures, the possible influence of long range pretransitional glass transition effects has to be considered. It is indeed proposed [6] that the elastic plateau is due to long range density fluctuations which are associated with the glass transition and which are frozen at the frequencies of the experiment. This approach shows that polymeric system contains intrinsically the glass properties which will produce, by decreasing the temperature, the non-equilibrium frozen state called glass transition.

We believe that this elastic property is probably shared by various viscoelastic materials and can certainly be enhanced by playing with the molecular architecture. The polymer melt dynamics is undoubtedly a challenging task.

*References*

1. C. Pujolle-Robic, L. Noirez, Nature **409,** 167 (2001); C. Pujolle-Robic, P.D. Olsmted, L. Noirez, Europhys. Lett. **59,** 364 (2002).

2. J.D. Ferry, Viscoelastic Properties of Polymers, 3$^{rd}$ ed, Wiley: New York 1980.

3. S. Hess, Z. Naturforsch **31**a, 1507 (1976); P.D. Olmsted, P. Goldbart, Phys. Rev. A**41**, 4578 (1990); ibid A**46,** 4966 (1992).

4. H.-W. Hu, G.A. Carson, S. Granick, Phys.Rev.Lett. **66,** 2758 (1991); H.-W. Hu, S. Granick, Science **258,** 1339 (1992); S. Granick, H.-W. Hu, Langmuir **10,** 3857 (1994); G.Reiter, A.Levent Demirel, S.Granick, Science, **263,** 1741 (1994), A. Levent Demirel & S. Granick, J. Am. Chem. Physics **115,** 1498 (2001); Y. Zhu, S. Granick, Phys. Rev. Lett. **93,** 096101 (2004);

5. J.L. Gallani et al., Phys. Rev. Lett. **72** 2109 (1994); P. Martinoty et al. Macromol. **32,** 1746 (1999);

6. D. Collin et al., Physica A. **320**, 235 (2002), See also note added in proof of Macromol. **32,** 1746 (1999) which indicates that similar elastic behaviours is identified in polysiloxanes substitued by long (15 carbons) side alkyl branches.

7. U. Tracht, et al., Phys. Rev. Lett. **81**, 2727 (1998)

8. E.W. Fischer, Physica A **201**, 183 (1993) ; E.W. Fischer et al., Acta Polym. **45,** 137 (1994) M. Beiner, et al., J. Non-Cryst. Solids **307**, 658 (2002) ; E.W. Fischer, et al., J. Non-Cryst. Solids **307**, 584 (2002); R. Walkenhorst, J. Chem. Phys. **109**, 11043 (1998).

9. H.R. Brand, K. Kawasaki, Physica A **324**, 484 (2003).

10. A.S. Bakai, J. Non-Cryst. Solids **307**, 623 (2002).

11. P.G. de Gennes & J. Prost , The Physics of Liquid Crystals (Oxford Science Publications, 1993); J. Bechhoefer, B. Jérôme, P. Pieransky, Phys. Rev.A **6,** 3187 (1990); B. Jérôme, J. Phys.: Condens. Matter **6,** A269 (1994); B. Jérôme, J. Commandeur, W.H. de Jeu, Liq. Cryst. **22,** 685 (1997).

12. N. Grassie et al., J. Polym. Sci., A-1 **9,** 931 (1971).

13. H. Mendil, L. Noirez, P. Baroni, submitted, L. Noirez to be published in Phys.Rev.E.

14. P. Baroni, H. Mendil, L. Noirez, patent submitted.

15. T. Young, Philos. Trans. R. Soc. London, **95** (1805) 65.

16. A.W. Adamson, Physical Chemistry of surfaces, 4th Ed; Wiley; New York (1982); P.G. de Gennes Rev. Mod. Phys. **57,** 827 (1985); P. Silberzan & L. Léger, Macromol. **25,** 1267 (1992).

17. H. Lakrout, C. Creton, D. Ahn, K. R. Shull, Macromol. **34**, 7448 (2001).

18. V. Fourmeaux-Demange et al, Macromol. **31,** 7445 (1998),

19. G. Reiter & R. Khanna, Phys. Rev. Lett. **85** 2753 (2000), Langmuir, **16** 6351 (2000).

20. J.F. Douglas, J.B. Hubbard, Macromol. **24,** 3163 (1991).

**Figures captions**

Fig.1a: Frequency dependence of the elastic modulus $G_1'$ of the LC-polymer (PAOCH$_3$) at different strain amplitudes (T=130°C (T-T$_{NI}$=+14°C), 0.300mm gap thickness, plate-plate surface treated fixtures (12 mm diameter)).

$\gamma$: (♦)0.3%(△)0.5%(▼)1%(□)2,5%(▲)5%(◇)10%(●)25%(▽)50%.

Fig.1b: Corresponding frequency dependence of the viscous modulus $G''$ versus strain amplitude:

$\gamma$: (□)2,5%(▲)5%(◇)10%(●)25%(▽)50%.

Fig.1c: Evolution of $G_p'$ of the LC-polymer (PAOCH$_3$) versus strain rate at different thicknesses at T=130°C (T-T$_{NI}$=+14°C): (■)0.025mm, (□)0.050mm, (▲)0.075mm, (△)0.100mm, (♦)0.150mm, (○)0.300mm. The inset displays the evolution of $G_p'$ as a function of the thickness at 130°C at $\gamma$=0.5%, 1% and 2.5% (log-log representation). The dotted lines are eye guides.

Fig.1d: Evolution of $\sigma = G_p'.\gamma$ versus strain rate of the LC-polymer (PAOCH$_3$) at different thicknesses at T=130°C (T-T$_{NI}$=+14°C).

(■)0.025mm(□)0.050mm(▲)0.075mm(△)0.100mm(♦)0.150mm(○)0.300mm.

Fig.1e: Evolution of the elastic plateau $G_p'$ of the LC-polymer (PAOCH$_3$) versus temperature ($\gamma$=0.5%, 0.300mm gap thickness).

Fig.2a: Evolution of the contact angle versus time of a Poly(n-butyl acrylate) (Table 1) in the molten state on different surfaces (○) aluminium (▲) treated surfaces. The photos at the left side represent a drop at t=0 and at the right side at t=70mn in the spreading state. In the inset is illustrated a schematic representation of a macroscopic drop. θ is the dynamic contact angle.

Fig.2b: Frequency dependence of $G'$(▲) and $G''$(○) measured on PBuA (Table 1) at $\gamma$=1% after prolonged time thermalisation (T=25°C, 0.060mm gap thickness, plate-plate surface treated fixtures (10 mm diameter)). The low modulus curve $G'$(▲) corresponds to the response of the transducer without sample.

Fig.2c: Master curve of the storage and loss moduli $G'$(▲) and $G''$(○) respectively measured on PBuA (Table1); $T_0=T_{ref}$=+25°C, $a_{T/T0}$=400. The dotted area indicates the frequency domain covered in Fig.2b.

Fig.2d: Frequency dependence of the viscoelastic moduli of the PBuA at different strain amplitudes (room temperature, 0.060mm gap thickness, plate-plate treated surface fixtures (10mm diameter)):

$G'(\omega)$:(●)0,5%(▷)1%(▼)2%(△)3%(▲)5%(◇)10%(■)20%(⊿)30%(♦)50%(▽)150%.

$G''(\omega)$:(▲)5%(◇)10%(■)20%.

Fig.2e: Evolution of the elastic plateau $G_p'$ of the PBuA, versus strain rate at different thicknesses at T=25°C (T-$T_g$=+89°C): (■)0.020mm(□)0.030mm(▲)0.040mm(△)0.060mm(♦)0.075mm(○)0.100mm. The inset displays the evolution of $G_p'$ as a function of the thickness at 25°C at $\gamma$=0.2%, 0.3% and 0.5% (log-log representation).

Fig.2f: Evolution of $\sigma=G_p'.\gamma$ versus strain rate of the PBuA at different thicknesses at T=25°C (T-$T_g$= +90°C): (■)0.020mm(□)0.030mm(▲)0.040mm(△)0.060mm(♦)0.075mm(○)0.100mm.

**Figures**

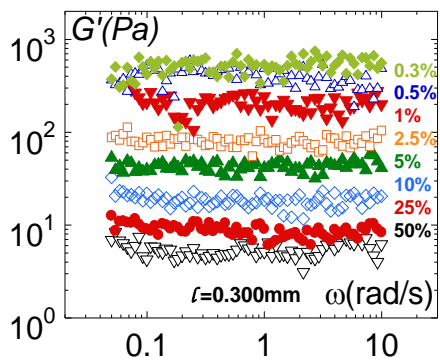

Fig.1a

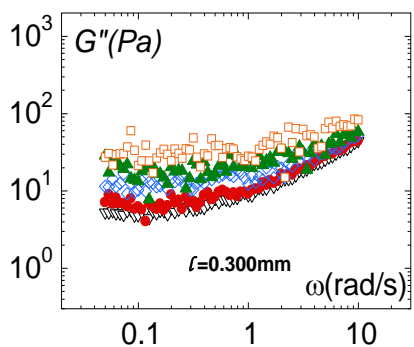

Fig.1b

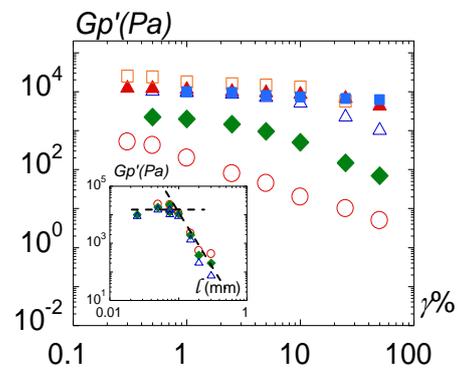

Fig.1c

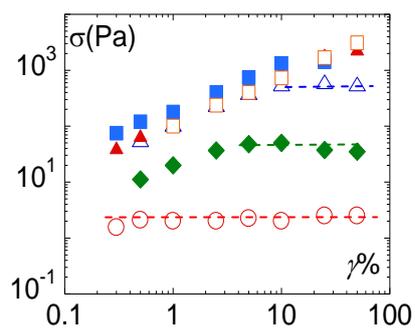

Fig.1d

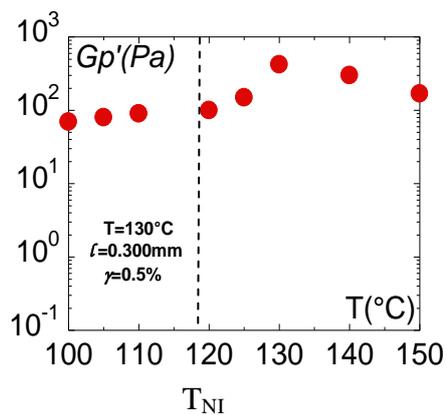

Fig.1e

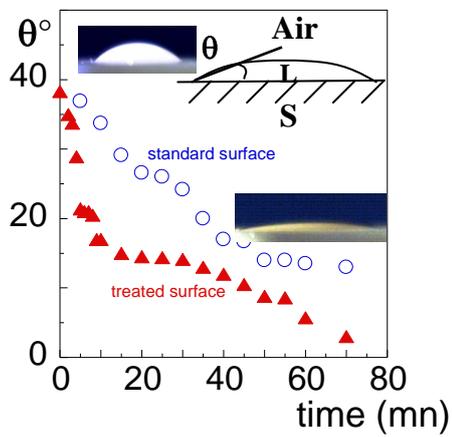

Fig.2a

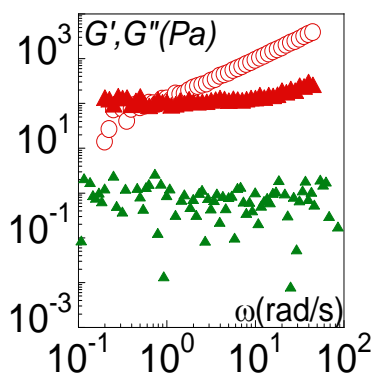

Fig.2b

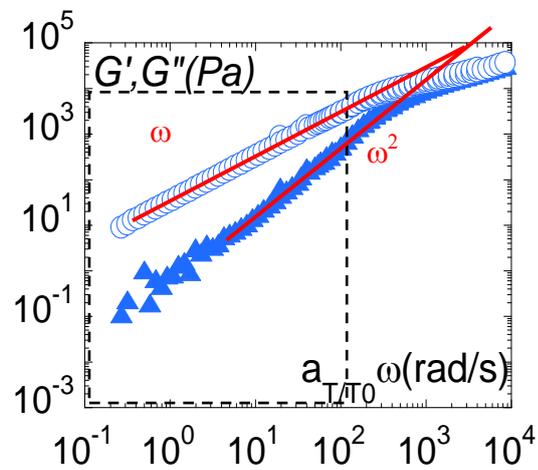

Fig.2.c

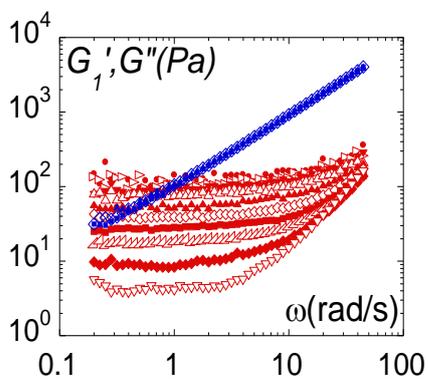

Fig.2d

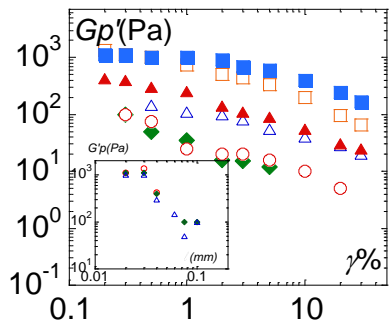

Fig.2e

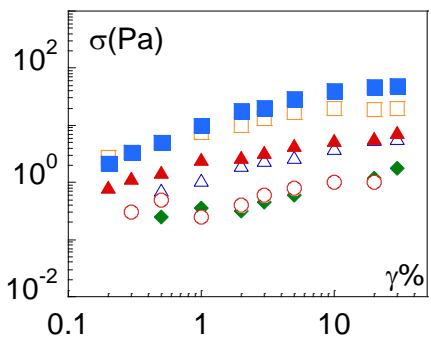

Fig.2f

16